\documentclass[10pt,noshowpacs,aps,twocolumn]{revtex4}
\usepackage{graphicx}% Include figure files
\begin{document}
%%{{{ Macroses
\newcommand{\Fig}{Figure~} 
\newcommand{\Figs}{Figures~}
\newcommand{\Figfirst}{Figure~}
\newcommand{\Rb}{${}^{87}$Rb }
\newcommand{\be}{\begin{equation}}
\newcommand{\ee}{\end{equation}}
\newcommand{\fxm}[1]{{\bf !---!FIXME #1 !---!}}
\newcommand{\etal}{\textit{et al.~}}
%%% %}}}

%{{{ authors
\author{Eugeniy E.\ Mikhailov}
	\email{
		evmik@tamu.edu,
		telephone: (979) 847-8593,
		fax: (979) 845-2590,
	}
	%\homepage{http://leona.physics.tamu.edu/~evmik/}
\author{Yuri V.\ Rostovtsev}
	\email{rost@atlantic.tamu.edu}
\author{George R.\ Welch}
	\email{grw@tamu.edu}

\affiliation{
	        Department of Physics,
		Texas A\&M University,
		College Station, Texas 77843-4242
}
%\thanks is used only if any from above is not applied
%\thanks{}
%}}}

\title{ %{{{
	Group velocity study in hot \Rb vapor with buffer gas
%	buffered media in EIT regime.  
%	\\or \\
%	Slowing with moving atoms in EIT regime.  
%	\\or \\
%	\dots
} %}}}

\begin{abstract} %{{{
	We study the behavior of the group velocity of light
	under conditions of electromagnetically induced
	transparency (EIT) in a Doppler broadened medium.
	Specifically, we show how the group delay (or group
	velocity) of probe and generated Stokes fields depends
	on the one-photon detuning of drive and probe fields.
	We find that for atoms in a buffer gas the group
	velocity decreases with positive one-photon detuning
	of the drive fields, and increases when the fields
	are red detuned.  This dependence is counter-intuitive
	to what would be expected if the one-photon detuning
	resulted in an interaction of the light with the
	resonant velocity subgroup.
%
%	????  What does this mean:
%	For the same experimental condition rise of density
%	atom decrease group velocity of probe field, when the
%	same rise increase group velocity of new field for
%	large enough densities.
%
%
\end{abstract} %}}} 

\date{\today}
\maketitle

\section{Introduction.} %{{{

	Electromagnetically induced transparency (EIT) has
been the subject of numerous theoretical and experimental
studies~\cite{brandt'97, vanier, helm'01, lukin97prl,
akulshin'98, akulshin'91, javan'02, yudin'00jl}.  Materials
displaying EIT have many interesting properties, such as
narrow resonance width and steep dispersion resulting in
ultra-low light group velocity~\cite{hau99, kash99, budker99}.
These properties make EIT interesting for several practical
applications, such as frequency standards~\cite{knappe2001},
precision magnetometry~\cite{novikova'01ol}, enhanced nonlinear
optics~\cite{hakuta91, hemmer95, jain96, zibrov99}, and
quantum information storage~\cite{phillips01prl, hau01nature,
zibrov02prl}.

	The group velocity of light in media can be expressed
as~\cite{Kochar2001prl}
\begin{equation}
v_g = {\frac{d \omega}{d k}} 
    = {
	    \frac{
	    c - \omega \frac{\partial n(\omega,k)} {\partial k} }
            {n(\omega,k) + \omega \frac {\partial n(\omega,k)} {\partial \omega}
       } }
    = \tilde{v}_g - v_s   
\end{equation}
where $c$ is the speed of light in vacuum, $\omega$ is the
frequency of the field, $n$ is the index of refraction in the
medium, $k$ is the vacuum wave number, and we have defined
\begin{equation}
\tilde{v}_g = 
    \frac{ c  } 
    {n(\omega,k) + \omega \frac {\partial n(\omega,k)} {\partial \omega} }
\end{equation}
and
\begin{equation}
v_s= 
    \frac{ \omega \frac{\partial n(\omega,k)} {\partial k} }
    {n(\omega,k) + \omega \frac {\partial n(\omega,k)} {\partial \omega} }~.
\end{equation}
Here and below we use the convention that terms with a tilde
(\,$\tilde{}$\,) denote values in the moving reference frame and
terms with no tilde denote values in the laboratory frame.

	The first term, $\tilde{v}_g$\,, is due to frequency
dispersion and the second term, $v_s$\,, is due to spatial
dispersion.  Because of spatial dispersion, the group velocity
is different for atoms with different speeds $v_a$\,.  This is
easily seen because in the moving frame $\tilde{\omega}={\omega}
- k v_a$ (we take $v_a$ positive for an atom moving in the
same direction as the light propagation), from which we see
that $v_g = {\frac{d \omega}{d k}} = \tilde{v}_g + v_a$.
This is just just the Galilean transformation from the
moving frame of the atoms to the laboratory frame.  Thus,
a mono-velocity atomic beam moving in the opposite direction
from the light propagation direction slows the group velocity.

	In a dense Doppler broadened medium, it
is possible to obtain slower or even zero group
velocity~\cite{Kochar2001prl,rostovtsev2002jmo} for the
probe field in a $\Lambda$ configuration of strong drive and
weak probe fields when both are red detuned but maintain the
two-photon resonance condition.  The following conditions on
the power of the drive field must be satisfied
\begin{equation}
	\label{eit_cond}
	\Omega \gg \sqrt{\gamma \gamma_{cb}} 
\end{equation}
\begin{equation}
	\label{quazi_beam_cond}
	\Omega  \ll k_d v_T \sqrt{ \frac{\gamma_{cb}} {\gamma} }  
\end{equation}
where $\Omega$ is the Rabi frequency of the drive laser and
$k_d$ is its wave vector, $v_{T}$ is the average thermal
velocity, $\gamma$ is the decay rate of the upper level, and
$\gamma_{cb}$ is the decay rate of coherence between the lower
(ground) levels.

	Equation~(\ref{eit_cond}) is just the usual condition
for EIT for individual atoms.  Equation~(\ref{quazi_beam_cond})
is applicable only for the case where the drive field
is weak enough that EIT occurs only for a narrow spread
of atomic velocities.  In this case, the intensity of
the drive field is not large enough to pump all atomic
velocity subgroups into the dark state.  This means that
the optical pumping rate ${|\Omega|^2\gamma/\Delta^2}$,
for atoms having one-photon detuning $\Delta$, is less
than the relaxation rate $\gamma_{bc}$, between levels $b$
and $c$.  Therefore ${|\Omega|^2\gamma/\Delta^2} <\gamma_{bc}$
implies that EIT does not occur for all moving atoms and
the light interacts with a quasi atomic beam.  When these
conditions are satisfied, one can choose a velocity sub-group
of atoms with a particular average velocity (in the direction
of light propagation) and narrow velocity spread $\Delta
v \approx \frac{\Omega}{k_d}\sqrt{\gamma_{cb}/\gamma}$.
This is accomplished by changing the one-photon detuning of
the drive laser field while maintaining two-photon resonance.
In this case, the center of the quasi-beam of moving atoms is
determined by the simple Doppler relation
\begin{equation}
\label{vel_of_center_beam}
v_a = c \frac{\Delta}{\omega_d}
\end{equation}   
where $\omega_d$ is the drive laser frequency.  Naturally
we would expect atoms moving with the atomic quasi-beam to
increase the group velocity, and atoms moving in the opposite
direction to ``drag'' light with them, or decrease the total
group velocity.

	The intuitive picture described above is quantified
rigorously in Ref.~\cite{Kochar2001prl}.  In the present paper,
we experimentally study these theoretical predictions by
measuring the dependence of the group velocity on probe-field
one-photon detuning for different experimental conditions.

	A common method of increasing the dispersion in an EIT
medium is to lengthen the ground-state coherence lifetime,
thereby decreasing the linewidth of the EIT resonance.
The coherence lifetime is often limited by the interaction time
of the atoms with the lasers.  Common methods for increasing
this lifetime are by introducing a buffer gas to confine
the atoms~\cite{brandt'97,wynands'98,helm'00,helm'01},
using wall coatings in the cell so that coherence is
preserved between successive interactions of the atoms and the
lasers~\cite{kanorsky'95, budker'98, budker'99ajp, robinson'58,
bouichiat'66, alexandrov'02prl}, and by cooling and trapping
the atoms~\cite{lazema'01optcom, marangos'98pra}.

	We have found that for EIT conditions in a sample with
buffer gas (with linewidth on the order of several kHz and group
velocity on the order of a few tens to hundreds of meters/sec)
the probe field has a slower group velocity when it is blue
detuned with respect to resonance and higher group velocity
for red detuning.  This result is opposite to the intuitive
picture described above, and to that of Kocharovskaya \etal
in \cite{Kochar2001prl}.

%}}} 

\section{Experiment} %{{{

\subsection{Setup} %{{{

\begin{figure}
\includegraphics[angle=0, width=1.00\columnwidth]{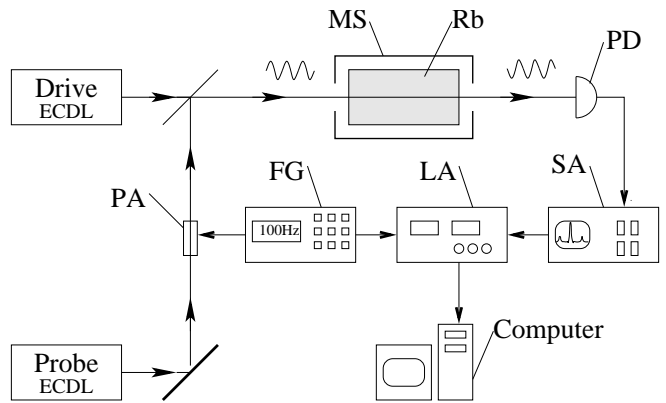}
\caption{
	\label{setup}
	Schematic of the experimental setup.
	MS is magnetic shield, 
	PD is fast photo diode,
	PA is power attenuator,
	FG is frequency generator,
	LA is lock-in amplifier,
	SA is spectrum analyzer.  
}
\end{figure}

	A schematic of the experimental setup is shown in
\Fig\ref{setup}.  One external cavity diode laser (ECDL) is
used as the source of a strong driving field and another as a
weak probe.  The drive and probe lasers are combined and pass
through a cell containing isotopically enhanced \Rb vapor.
The \Rb density is varied by changing the temperature of
the cell.  A three layer magnetic shield (MS) screens out the
laboratory magnetic field.

\begin{figure}
\includegraphics[angle=0, width=0.75\columnwidth]{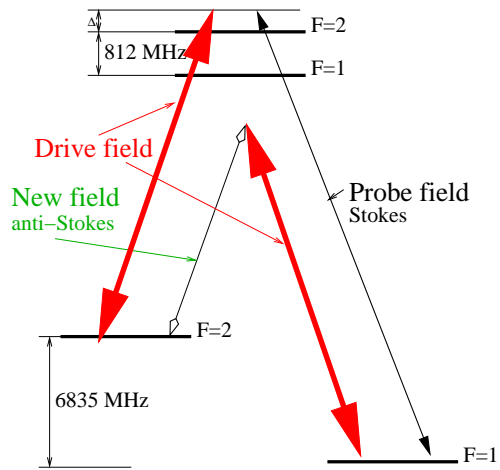}
\caption{
	\label{levels}
	Relevant levels in $^{87}$Rb.
}
\end{figure}

	The lasers are phase locked to each other with a
frequency offset that is tunable about the ground level
hyperfine splitting of $^{87}$Rb (6835 MHz).  The lasers
are spatially mode-matched with a single-mode optical
fiber.  The drive laser is tuned to the $5S_{1/2}(F=2) \to
5P_{1/2}(F'=2)$ transition of \Rb and the probe laser is tuned
to the $5S_{1/2}(F=1) \to 5P_{1/2}(F'=2)$ transition as shown
in \Fig\ref{levels}.

	The configuration of drive and probe lasers shown
in \Fig\ref{levels} is called a $\Lambda$ configuration.
In this case, the lasers optically pump all atoms in the
desired velocity subgroup into a dark state superposition of
the ground levels, giving rise to strong coherence between
these lower levels.  Scattering of the drive field on this
coherence results in the generation of a fairly strong Stokes
component (new field) in the medium (shown in \Fig\ref{levels}).
The generation of this new field near two-photon resonance of
the drive and probe field is described in Ref.~\cite{lukin97prl,
kash99, mikhailov2002}.

	The transmitted drive and probe fields and the generated
Stokes field are detected separately by heterodyne detection on
a fast photo detector.  This is done by splitting off part of
the drive field from the main beam before entering the cell.
This component is shifted up in frequency by a small amount
(60~MHz), and combined with the beams exiting the cell before
the photo-detector.  The photo-current thus contains beat
signals at various RF frequencies, separated by 60~MHz, in the
vicinity of the 6835~MHz separation of the drive and probe.
By separately analyzing these components with a spectrum
analyzer (SA), we extract the transmitted probe and generated
new fields independently.  This technique is described in
Ref.~\cite{mikhailov2002,kash99}.

%}}}

\subsection{Time delay and group velocity measurement procedure.} %{{{

	We extract the group velocity in the medium by
modulating the intensity of the probe field before the cell,
and observing the time delay before this modulation is observed
in the transmitted field.  Experiments have been conducted
with gaussian shaped (temporal) pulses and with sinusoidal
modulation.  We find that the delay time is independent of the
probe field modulation technique as long as the bandwidth of
the modulation does not exceed the transmission linewidth of
the EIT resonance.

	The modulation is generated by use of a frequency
generator (FG) which drives an acousto-optical (AO) modulator
in the probe laser.  The deflected beam from the AO is
blocked, so the AO serves as a probe power attenuator (PA
in \Fig\ref{setup}).  We also measure the time delay of the
generated new field.

	With a sinusoidal modulation of the probe, lock-in
detection of the transmitted probe field provides a sensitive
measure of the time delay due to the medium.  When using lock-in
detection in this way, we obtain the time delay from the phase
shift of the transmitted probe field intensity relative to the
probe intensity before the cell.  For a sine wave of frequency
$f$, this phase shift is given by
\begin{equation}
\psi = 2\pi\tau_{d}f-\psi_{0} \,,
\end{equation}
where $\tau_{d}$ is the time delay introduced by the atoms
and $\psi_{0}$ is a phase shift introduced by electronics.
To eliminate the unknown $\psi_{0}$ we measure the phase shift
for several different modulation frequencies.  The phase
shift increases linearly with frequency $f$ and the slope
of this line is $2\pi\tau_d$.  We extract $\tau_d$ with a
least squares fit.  We then find the group velocity $v_g$ by
setting $v_g = L/\tau_{d}$ where $L$ is the length of the cell.
($L\approx 1~\mathrm{cm}$ for our experiment.)

	Another experimental technique to simplify the group
velocity measurement is to replace the second laser and
phase-lock circuitry with an electro-optic modulator (EOM) in
the drive laser beam.  By applying a narrow tunable microwave
signal to the EOM, upper and lower sidebands are generated at
the microwave frequency, which we choose to match the 6.835~GHz
ground state hyperfine splitting.  We tune the laser to the
drive transition, so that the upper sideband drives the probe
transition and the lower sideband is off resonance.  We choose
the microwave amplitude to generate a sideband with power that
is 1/10 of the drive power.  A careful comparison of the two
methods (two phased locked lasers versus one modulated laser)
shows no difference in the group velocity measurements.
We note that a similar technique is to modulate the laser
current, which directly creates sidebands on the laser, has
also been successfully employed~\cite{affolderbach'99}.

	We measure the dependence of $v_g$ on one-photon laser
detuning $\Delta=\omega_d-\omega_{22}$\,, where $\omega_d$
is the frequency of the drive laser and $\omega_{22}$ is the
frequency of the $F=2 \to F'=2$ transition.  During each such
measurement, the frequency difference of the probe and drive
lasers is kept constant and equal to ground level splitting
(6.835 GHz).  The drive laser power is $300 \mu W$, and probe
power is $3 \mu W$.  These measurements may then be repeated for
different \Rb densities (different temperatures of the cell).

%}}}

\subsection{Experimental results} %{{{

\begin{figure}
\includegraphics[angle=0,
	width=1.00\columnwidth]
	{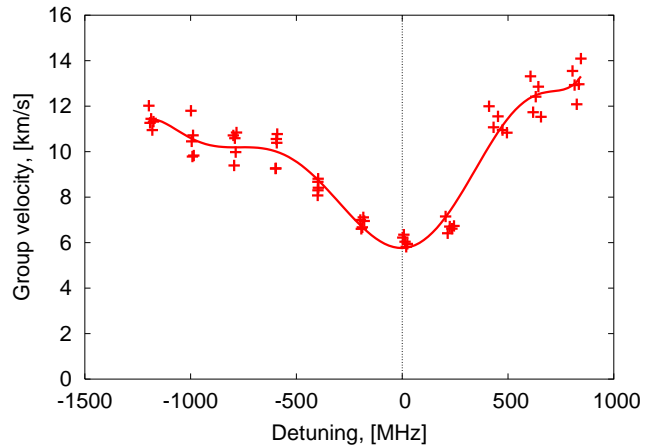}
\caption{
	\label{pr_vg_vs_D_in_vacuum}
	Group velocity vs detuning of \Rb atoms for probe
	field in cell with no buffer gas.  Cell length
	$L=47.5~\mathrm{mm}$ and the density is $3.6\times
	10^{11}~{\mathrm{cm}}^{-3}$.  Drive power input to the
	cell is $1310 \mu W$, and transmitted drive power is
	$741 \mu W$.  Data points are shown by a +.  The solid
	curve is a seventh-order polynomial fit to the data
	points and is shown only as a guide for the eye.
%
	%T=64.5 correspond to N=.36 10^{12}}
%
}
\end{figure}

	We first consider the case where no buffer gas is used,
and the ground-state coherence lifetime is limited by the
free-flight transit time of the thermal rubidium atoms through
the laser beam.  In our experiment, the EIT transmission
linewidth is 30~kHz  and the resulting group velocity on
the order of 10~km/s.  \Figfirst\ref{pr_vg_vs_D_in_vacuum}
shows the dependence of the group velocity as a function
of the drive laser frequency in the vicinity of the drive
resonance.  The group velocity is too large to observe the
spatial dispersion effect described in the introduction.
Another way to see this is to note that the group velocity
is much higher than the mean thermal speed of the atoms.
In \Fig\ref{pr_vg_vs_D_in_vacuum} the drive detuning spans
the full range of the upper-state hyperfine splitting, and
the small feature on the left of the spectrum is the result of
generating EIT on the upper $F'=1$ level (see \Fig\ref{levels}).
%
%{{{ 
%}}}

\begin{figure}
\includegraphics[angle=0,
	width=1.00\columnwidth]
	{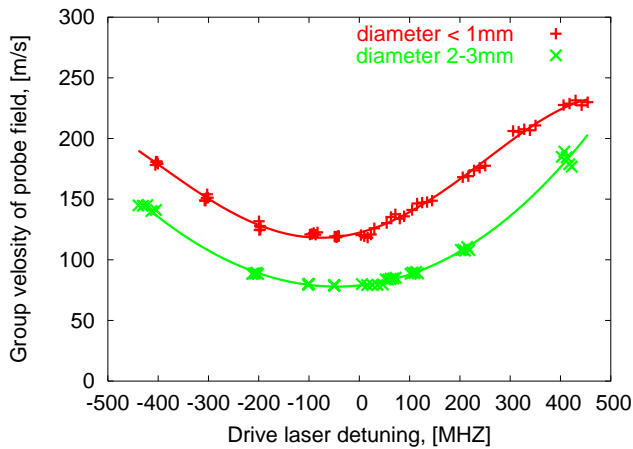}
\caption{
	\label{pr_vg_vs_D_and_beam_diameter_EOM}
	Probe field group velocity vs drive laser one-photon
	detuning in a cell with 3 torr of $\mathrm{N}_2$
	buffer gas.  Cell length $L=10~\mathrm{mm}$ and the
	density is  $8.7\times 10^{11}~{\mathrm{cm}}^{-3}$.
	The two curves are for beam diameters of 1 and 3
	mm respectively.
%
	%T=75.7 correspond to N=.87 10^{12}}
%
}
\end{figure}

	Next we narrow the transmission linewidth
with a buffer gas.  This increases the dispersion
considerably, resulting in reduced group velocity.
\Figfirst\ref{pr_vg_vs_D_and_beam_diameter_EOM} shows the
group velocity as a function of drive laser detuning for
similar conditions as \Fig\ref{pr_vg_vs_D_in_vacuum} but with
the addition of 3 torr of $\mathrm{N}_2$ buffer gas.  The EIT
transmission linewidth is only a few kHz and we see that the
group velocity has fallen to below 100~m/s.  We also see that
increasing the laser beam diameter increases the dispersion
and reduces the group velocity as expected.
%
%{{{ 
%}}}

\begin{figure}
\includegraphics[angle=0,
	width=1.00\columnwidth]
	{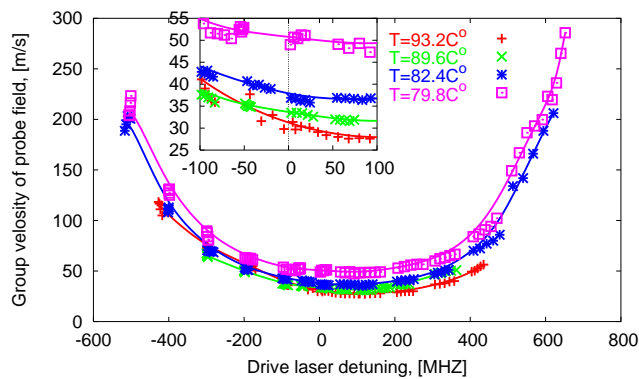}
\caption{
	\label{pr_vg_vs_detuning}
	Probe field group velocity vs drive laser
	one-photon detuning, for the same cell as in
	\Fig\ref{pr_vg_vs_D_and_beam_diameter_EOM}.  Drive power
	is $300 \mu W$ and probe power $3\mu W$.  
	%--------Referee is right this is confusing ---------%
	% Curves shown
	% are for successively increasing atomic density, shown
	%----------------------------------------------------%
	Curves are shown for different atomic density, measured
	in ${\mathrm{cm}}^{-3}$.  Inset: blow-up of the data
	in the vicinity of resonance, showing an increase to
	the red and decrease to the blue of resonance.
}
\end{figure}

	\Figfirst\ref{pr_vg_vs_detuning} shows how the group
velocity depends on drive laser detuning for increasing atomic
density.  We observe a lowering of the group velocity for
higher density, but in no case do we observe a lowering of the
group velocity as the drive laser is detuned red of resonance,
which we would expect based on the prediction of dragging
slow light by atoms moving opposite to the laser propagation
direction~\cite{Kochar2001prl}.  On the contrary, we see the
group velocity increase for negative drive laser detuning,
and a minimum group for one-photon detuning about 100~MHz blue
of one-photon resonance (see \Fig\ref{pr_vg_vs_detuning}).
%
%{{{ 
%}}}

	The explanation of this effect is still not clear,
but it is plausible that this counter-intuitive behavior is
caused by velocity changing collisions in the presence of the
buffer gas.  We note that when the beam diameter decreases,
the group velocity dependence versus one-photon detuning
becomes more shifted to the side of negative detuning (see
\Fig\ref{pr_vg_vs_D_and_beam_diameter_EOM}).  In other words,
the behavior becomes less counter-intuitive.  We explain
this by noting that for a very narrow beam, fewer velocity
changing collisions take place before the atom leaves the
laser beam.  In any case, it is clear that the discussion of
Ref.~\cite{Kochar2001prl} about a quasi mono-velocity beam
is not applicable in the case with a buffer gas, since all
velocity groups are mixed by velocity changing collisions.

	When the density of \Rb atoms is increased ($\approx
10^{12}~\mathrm{cm}^{-3}$) highly nonlinear interaction of the
drive and probe fields leads to very efficient generation of a
Stokes component, or new field~\cite{lukin97prl,kash99}.  We can
measure the intensity of the generated field as a function of
two-photon detuning and find the width to be greater than the
transmission width of the EIT resonance of the probe field.
(Under the conditions in our experiment it is roughly a factor
of two wider~\cite{mikhailov2002}.)  Correspondingly, we also
measure the time delay between modulation of the probe field
before the cell and the resulting modulation of the new field
after the cell.  For the rubidium cell with buffer gas, this
delay time is smaller than for the probe field, meaning that
the group velocity of the new field ($v_n$) is greater than
the group velocity of the probe field ($v_p$).  This is no
great surprise since the new field is propagating far from
one-photon resonance.

\begin{figure}
\includegraphics[angle=0,
	width=1.00\columnwidth]
	{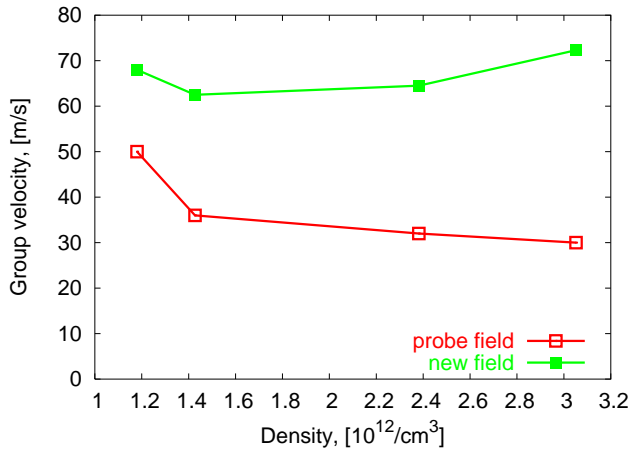}
\caption{
	\label{group_velosity_vs_density}
	Group velocity vs density of \Rb atoms for
	probe and new fields, for the same cell as
	in \Fig\ref{pr_vg_vs_D_and_beam_diameter_EOM}.
	Drive power is $300 \mu W$, and probe power is $3\mu W$.
}
\end{figure}

	As discussed above, as the rubidium density is increased
the probe field group velocity decreases.  Similar behavior
occurs for the generated new field for low density.  However,
for large atomic density the group velocity for the new field
starts to increase with density.  These results are shown
in \Fig\ref{group_velosity_vs_density}.  This dependence is
connected with a propagation effect.  The effective generation
of new field occurs in the part of the cell where the group
velocity is small.  As the light propagates through the cell,
the drive field intensity decreases until the new field is
decoupled from the probe.  After this point, the new field
propagates at nearly the vacuum speed of light. Thus the
observed average speed of new field increases with atomic
density.

%	As discussed above, as the rubidium density is increased
%the probe field group velocity decreases.  Similar behavior
%occurs for the generated new field for low density.  However,
%for large atomic density the group velocity for the new field
%starts to increase with density.  
%These results are shown in \Fig\ref{group_velosity_vs_density}.
%
% This can be explained by considering that in dense media the
% width of the EIT resonance decreases with increasing density of
% \Rb \cite{sautenkov99las, lukin97prl}
% thus $v_g$ decreases too because
%
% \begin{equation}
% \label{group_vel_vs_width}
% v_g \sim \frac{8 \pi }{3 \lambda^2 N}\gamma_r  
% \end{equation}
%
% where $\gamma_r$ is the width of the resonance, $N$ density of \Rb 
% atoms, $\lambda$ wave length of the probe field .  
%
%
%{{{
%}}}
%

%}}}

\begin{figure}
\includegraphics[angle=0,
	width=1.00\columnwidth]
	{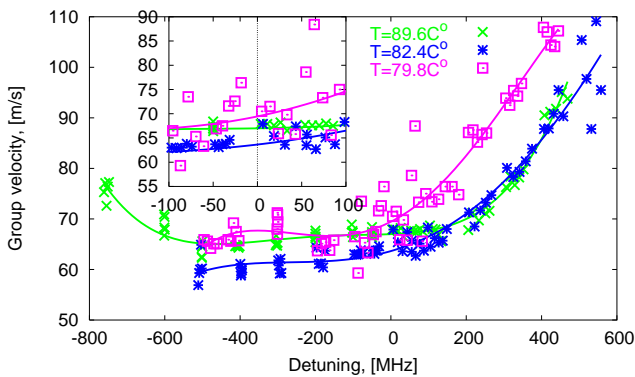}
\caption{
	\label{nw_vg_vs_detuning}
	New field group velocity vs drive laser
	one-photon detuning, for the same cell as
	in \Fig\ref{pr_vg_vs_D_and_beam_diameter_EOM}.
	Drive power is $300 \mu W$ and probe power $3\mu W$.
	%--------Referee is right this is confusing ---------%
	% Curves shown
	% are for successively increasing atomic density, shown
	%----------------------------------------------------%
	Curves are shown for different atomic density, measured
	in ${\mathrm{cm}}^{-3}$.  Inset: blow-up of the data
	in the vicinity of resonance, showing an increase to
	the red and decrease to the blue of resonance.
}
\end{figure}

	We can also measure the new-field group velocity as
a function of drive laser detuning.  The results are shown
in \Fig\ref{nw_vg_vs_detuning}.  We find that the the group
velocity is smaller for negative one-photon detuning than for
positive detuning in vicinity of resonance.  This behavior
follows the intuitive predictions of Ref.~\cite{Kochar2001prl}.
%
%{{{ 
%}}}

%%%%{{{ 
%%%\begin{figure}
%%%\includegraphics[angle=0,
%%%	width=1.00\columnwidth]
%%%	{experim_results/sep25_2-3mm_wide_beam_EOM/probe_2-3mm_wide_beam}
%%%\caption{
%%%	\label{pr_vg_vs_D_EOM}
%%%%
%%%	Probe field group velocity with drive laser field one
%%%	photon detuning in 1cm long cell with 3 torr of $\mathrm{N}_2$
%%%	as a buffer gas.  Diameter of the beam 1-2 mm . Setup
%%%	with EOM.  Temperature is  $93.2^oC$
%%%%
%%%}
%%%\end{figure}
%%%%}}}

%}}}

%}}}

%%%%%%%%%%%%%%%%%%%%%%%%%%%%%%%%%%%%%%%%%%%%%%%%%%%%%%%%%%%%%%%%%%%%%%%%%%%
%\section{Theory.} %{{{
%\subsection{Simplified equations}
%\fxm{ It would be nice to have some simplified scheme consideration here}
%
%\subsection{Theoretical results}
%\fxm{	 Something here too}
%}}}

%%%%%%%%%%%%%%%%%%%%%%%%%%%%%%%%%%%%%%%%%%%%%%%%%%%%%%%%%%%%%%%%%%%%%%%%%%%
\section{Summary.}

	We observe counter-intuitive dependence of
the probe field group velocity versus drive field one-photon
detuning for different densities (temperatures) of \Rb (see
\Fig\ref{pr_vg_vs_detuning}).  The group velocity decreases
slightly for blue-detuned drive fields and increases slightly
for red-detuned fields.  We conclude that the the predictions
of Ref.~\cite{Kochar2001prl} cannot be applied to the case
where the EIT linewidth is reduced with a buffer gas, since
all velocities are constantly mixing via velocity changing
collisions.  Unfortunately, without a buffer gas we cannot
achieve narrow enough EIT resonances to reach the amount of
dispersion needed to get group velocities low enough to observe
the effect of dragging the light by atoms.

	We find that the group velocity is higher for the
generated Stokes field, and that the behavior as a function
of detuning is opposite that of the probe field.

\section{Acknowledgement} 

	We thank A.\ B.\ Matsko, I. Novikova, V. Sautenkov,
and M.\ O.\ Scully for useful and stimulating discussions.
This work was supported by the Office of Naval Research.

%%%%%%%%%%%%%%%%%%%%%%%%%%%%%%%%%%%%%%%%%%%%%%%%%%%%%%%%%%%%%%%%%%%%%%%%%%%

\bibliographystyle{journ_of_modern_optics}

\begin{thebibliography}{}

\bibitem[Brandt et~al., 1997]{brandt'97}
Brandt, S., Nagel, A., Wynands, R., and Meschede, D., 1997, {\em Phys. Rev. A},
  {\bf 56}(2), R1063--R1066.

\bibitem[Vanier et~al., 1998]{vanier}
Vanier, J., Godone, A., and Levi, F., 1998, {\em Phys. Rev. A}, {\bf 58}(3),
  2345--2358.

\bibitem[Erhard and Helm, 2001]{helm'01}
Erhard, M. and Helm, H., 2001, {\em Phys. Rev. A}, {\bf 63}(4), 043813--+.

\bibitem[Lukin et~al., 1997]{lukin97prl}
Lukin, M.~D., Fleischhauer, M., Zibrov, A.~S., Robinson, H.~G., Velichansky,
  V.~L., Hollberg, L., and Scully, M.~O., 1997, {\em Phys. Rev. Lett.}, {\bf
  79}(16), 2959--2962.

\bibitem[Akulshin et~al., 1998]{akulshin'98}
Akulshin, A.~M., Barreiro, S., and Lezama, A., 1998, {\em Phys.\ Rev.\ A}, {\bf
  57}(4), 2996 --3002.

\bibitem[Akulshin et~al., 1991]{akulshin'91}
Akulshin, A.~M., Celikov, A.~A., and Velichansky, V.~L., 1991, {\em Opt.
  Commun.}, {\bf 84}(3-4), 139--143.

\bibitem[Javan et~al., 2002]{javan'02}
Javan, A., Kocharovskaya, O., Lee, H., and Scully, M.~O., 2002, {\em Phys. Rev.
  A}, {\bf 66}(1), 013805--+.

\bibitem[Taichenachev et~al., 2000]{yudin'00jl}
Taichenachev, A.~V., Tumaikin, A.~M., and Yudin, V.~I., 2000, {\em Jetp Lett.},
  {\bf 72}(3), 119--122.

\bibitem[Hau et~al., 1999]{hau99}
Hau, L.~V., Harris, S.~E., Dutton, Z., and Behroozi, C.~H., 1999, {\em Nature},
  {\bf 397}(6720), 594--598.

\bibitem[Kash et~al., 1999]{kash99}
Kash, M.~M., Sautenkov, V.~A., Zibrov, A.~S., Hollberg, L., Welch, G.~R.,
  Lukin, M.~D., Rostovtsev, Y., Fry, E.~S., and Scully, M.~O., 1999, {\em Phys.
  Rev. Lett.}, {\bf 82}(26), 5229--5232.

\bibitem[Budker et~al., 1999a]{budker99}
Budker, D., Kimball, D.~F., Rochester, S.~M., and Yashchuk, V.~V., 1999a, {\em
  Phys. Rev. Lett.}, {\bf 83}(9), 1767--1770.

\bibitem[Knappe et~al., 2001]{knappe2001}
Knappe, S., Wynands, R., Kitching, J., Robinson, H.~G., and Hollberg, L., 2001,
  {\em J. Opt. Soc. Am. B-Opt. Phys.}, {\bf 18}(11), 1545--1553.

\bibitem[Novikova et~al., 2001]{novikova'01ol}
Novikova, I., Matsko, A.~B., and Welch, G.~R., 2001, {\em Opt. Lett.}, {\bf
  26}, 1016--1018.

\bibitem[Hakuta et~al., 1991]{hakuta91}
Hakuta, K., Marmet, L., and Stoicheff, B.~P., 1991, {\em Phys. Rev. Lett.},
  {\bf 66}(5), 596--599.

\bibitem[Hemmer et~al., 1995]{hemmer95}
Hemmer, P.~R., Katz, D.~P., Donoghue, J., Croningolomb, M., Shahriar, M.~S.,
  and Kumar, P., 1995, {\em Opt. Lett.}, {\bf 20}(9), 982--984.

\bibitem[Jain et~al., 1996]{jain96}
Jain, M., Xia, H., Yin, G.~Y., Merriam, A.~J., and Harris, S.~E., 1996, {\em
  Phys. Rev. Lett.}, {\bf 77}(21), 4326--4329.

\bibitem[Zibrov et~al., 1999]{zibrov99}
Zibrov, A.~S., Lukin, M.~D., and Scully, M.~O., 1999, {\em Phys. Rev. Lett.},
  {\bf 83}(20), 4049--4052.

\bibitem[Phillips et~al., 2001]{phillips01prl}
Phillips, D.~F., Fleischhauer, A., Mair, A., Walsworth, R.~L., and Lukin,
  M.~D., 2001, {\em Phys. Rev. Lett.}, {\bf 86}(5), 783--786.

\bibitem[Liu et~al., 2001]{hau01nature}
Liu, C., Dutton, Z., Behroozi, C.~H., and Hau, L.~V., 2001, {\em Nature}, {\bf
  409}, 490--493.

\bibitem[Zibrov et~al., 2002]{zibrov02prl}
Zibrov, A.~S., Matsko, A.~B., Kocharovskaya, O., Rostovtsev, Y.~V., Welch,
  G.~R., and Scully, M.~O., 2002, {\em Phys. Rev. Lett.}, {\bf 88}(10), 103601.

\bibitem[Kocharovskaya et~al., 2001]{Kochar2001prl}
Kocharovskaya, O., Rostovtsev, Y., and Scully, M.~O., 2001, {\em Phys. Rev.
  Lett.}, {\bf 86}(4), 628--631.

\bibitem[Rostovtsev et~al., 2002]{rostovtsev2002jmo}
Rostovtsev, Y.~V., Kocharovskaya, O., and Scully, M.~O., 2002, {\em Journal of
  Modern Optics}, {\bf 49}(14/15), 2637--2643.

\bibitem[Wynands and Nagel, 1998]{wynands'98}
Wynands, R. and Nagel, A., 1998, {\em Appl.\ Phys.\ B}, {\bf 68}, 1 --+.

\bibitem[Erhard et~al., 2000]{helm'00}
Erhard, M., Nu{\ss}mann, S., and Helm, H., 2000, {\em Phys. Rev. A}, {\bf 62},
  061802(R)--+.

\bibitem[Kanorsky et~al., 1995]{kanorsky'95}
Kanorsky, S.~I., Weis, A., and Skalla, J., 1995, {\em Appl.\ Phys.\ B}, {\bf
  60}, S165 --+.

\bibitem[Budker et~al., 1998]{budker'98}
Budker, D., Yashchuk, V., and Zolotorev, M., 1998, {\em Phys.\ Rev.\ Lett.},
  {\bf 81}, 5788 --+.

\bibitem[Budker et~al., 1999b]{budker'99ajp}
Budker, D., Orlando, D.~J., and Yashchuk, V., 1999b, {\em Am.\ J.\ Phys.}, {\bf
  67}, 584 --+.

\bibitem[Robinson et~al., 1958]{robinson'58}
Robinson, H.~G., Ensberg, E.~S., and Dehmelt, H.~G., 1958, {\em Bull. Am. Phys.
  Soc.}, {\bf 3}, 9 --+.

\bibitem[Bouchiat and Brossel, 1966]{bouichiat'66}
Bouchiat, M.~A. and Brossel, J., 1966, {\em Phys.\ Rev.}, {\bf 147}, 41 --+.

\bibitem[Alexandrov et~al., 2002]{alexandrov'02prl}
Alexandrov, E.~B., Balabas, M.~V., Budker, D., English, D.~S., Kimball, D.~F.,
  Li, C.~H., and Yashchuk, V., 2002, {\em Phys. Rev. Lett}, {\bf 66}, 042903
  --+.

\bibitem[Lipsich et~al., 2001]{lazema'01optcom}
Lipsich, A., Barreiru, S., Valente, P., and Lezama, A., 2001, {\em Opt.\
  Commun.}, {\bf 190}(1-6), 185 -- 191.

\bibitem[Chen et~al., 1998]{marangos'98pra}
Chen, H.~X., Durrant, A.~V., Marangos, J.~P., and Vaccaro, J.~A., 1998, {\em
  Phys. Rev. A}, {\bf 58}(2), 1545 -- 1548.

\bibitem[Mikhailov et~al., 2002]{mikhailov2002}
Mikhailov, E.~E., Rostovtsev, Y., and Welch, G.~R., 2002, {\em Journal of
  Modern Optics}, {\bf 49}(14/15), 2535--2542.

\bibitem[Affolderbach et~al., 1999]{affolderbach'99}
Affolderbach, C., Nagel, A., Knappe, S., Jung, S., Wiedenmann, D., and Wynands,
  R., 1999, {\em Applied Physics B}, {\bf 70}(3), 407 -- 413.

\end{thebibliography}
%%%%%%%%%%%%%%%%%%%%%%%%%%%%%%%%%%%%%%%%%%%%%%%%%%%%%%%%%%%%%%%%%%%%%%%%%%%

% here we generatelist of figures %{{{  
	\newpage
% some how revtex have a broken defenition of \setminuscontentsline' comand
	\renewcommand{\contentsline}[3]{\flushleft Figure  #2 \\ at page #3 }
	\listoffigures
%}}}

\end{document}